\documentclass[12pt,preprint]{aastex}

\def\feka{Fe K$\alpha$}
\def\chandra{{\it Chandra}}
\def\xmm{{\it XMM-Newton}}
\def\asca{{\it ASCA}}

\def\sax{{\it BeppoSAX}}

\def\rosat{{\it ROSAT}}

\def\integral{{\it Integral}}

\def\glast{{\it GLAST}}
\def\swift{{\it Swift}}
\def\suzaku{{\it Suzaku}}

\def\lum{erg s$^{-1}$}
\def\flux{erg cm$^{-2}$ s$^{-1}$}   
\def\nh{cm$^{-2}$}
\def\arcsec{$^{\prime\prime}$}

\def\ltsima{$\; \buildrel < \over \sim \;$}
\def\simlt{\lower.5ex\hbox{\ltsima}} 
\def\gtsima{$\; \buildrel > \over \sim \;$}
\def\simgt{\lower.5ex\hbox{\gtsima}} 

\begin{document}
\title{\swift\ observations of high-redshift radio-loud quasars}


\author{R.M. Sambruna} 
\affil{NASA's GSFC, Code 661, Greenbelt, MD 20771}

\author{F. Tavecchio and G. Ghisellini}
\affil{Osservatorio Astronomico di Brera, via Brera 28, 20121 Milano,
Italy}

\author{D. Donato, S. T. Holland, C. B. Markwardt, J. Tueller, R. F. Mushotzky}
\affil{NASA's GSFC, Code 661, Greenbelt, MD 20771} 

\begin{abstract}

We report on \swift\ observations of four $z>2$ radio-loud quasars
(0212+735, 0537--286, 0836+710, and 2149--307), classified as
blazars. The sources, well-known emitters at soft-medium X-rays, were
detected at $>5\sigma$ with the BAT experiment in 15--150~keV. No flux
variability was detected within the XRT and BAT exposures, with the
exception of 0836+710 which shows an increase of a factor 4 of the
15--150~keV flux on a timescale of one month. The 0.3--10~keV spectra
are well fitted by power law models, with rather hard continua (photon
indices $\Gamma_{XRT} \sim 1.3-1.5$); similarly, the 15--150~keV
spectra are described by power laws with $\Gamma_{BAT} \sim 1.3-1.8$.
The XRT data exhibit spectral curvature, which can be modeled either
in terms of excess absorption along the line of sight, or a
downward-curved broken power law. In the former case, if the excess
N$_H$ is at the rest-frame of the source, columns of
N$_H^z=(0.3-6)\times 10^{22}$ \nh\ are measured. Modeling of the
SEDs of the four quasars shows that the emission at the higher
frequencies, \gtsima $10^{16}$ Hz, is dominated by the jet, while the
steep optical-to-UV continua, observed with the UVOT, can be
attributed to thermal emission from the accretion disk.  The disk
luminosity is between 1\% and 10\% the jet power, similar to other
powerful blazars.

\end{abstract}


\section{Introduction}

Blazars are radio-loud Active Galactic Nuclei (AGN) dominated by
non-thermal continuum emission from a relativistic jet closely aligned
to the line of sight. Blazars include the luminous Flat Spectrum Radio
Quasars (FSRQs), characterized by strong (EW $>$ 5\AA) broad optical
lines, and closer-by BL Lacs, where the optical lines are weak or
absent. The Spectral Energy Distributions (SEDs) of blazars, both
FSRQs and BL Lacs, span the whole range of the observed
electromagnetic spectrum, from radio to GeV and TeV gamma-rays (e.g.,
Ulrich, Maraschi, \& Urry 1997).  They are characterized by two broad
humps. The first one, ranging from radio to UV/X-ray wavelengths, is
widely attributed to synchrotron emission from the jet; the second
extends from X-rays to GeV and TeV gamma-rays, and its origin is less
well established. A popular interpretation is inverse Compton
scattering of ambient photons, either internal (synchrotron-self
Compton, SSC; Maraschi et al. 1992) or external (External Compton
scattering, EC; Dermer \& Schlickeiser 1993, Sikora et al. 1994) to
the jet.

In FSRQs the synchrotron component usually peaks at IR/optical
wavelengths. The hard ($\Gamma \sim 1.5$) \asca\ and \sax\ spectra,
together with the steeper EGRET spectra, imply a peak of the IC
component around a few GeV. While the low- and high-energy ends of the
IC component have been relatively well studied, little is known about
the emission in the intermediate hard X-ray/soft gamma-ray energy
range. The \sax\ PDS instrument detected a few FSRQs, including two of
the sources of this paper, up to $\sim$ 100~keV with a rather hard
continuum, photon index $\Gamma \sim 1.4$ (Tavecchio et al. 2000,
Elvis et al. 2003), while only a handful of sources were detected with
COMPTEL (Blom et al. 1996; Collmar 2006). More recently,
\integral\ detected a few blazars in outburst above 10~keV (3C~454.3
at $z$=0.86, Pian et al. 2006; NRAO~530 at $z$=0.9, Foschini et
al. 2006). With its broad-band coverage, \suzaku\ is allowing some of
the first investigations of the high-energy continuum emission from
blazars (RBS~315; Tavecchio et al. 2007).  

There are compelling reasons for studying the hard X-ray/soft
gamma-ray emission of blazars. First, this is the energy range that
connects the onset of the IC component at a few keV to the GeV
emission; thus its spectral quantification is important to
characterize the peak of the IC component and related physical
quantities (e.g., Ghisellini et al. 1998, Tavecchio et
al. 2000). Second, in the soft X-ray band a ``break'' is expected in
the EC component, tracking the low-energy tail end of the electron
energy distribution (Tavecchio et al. 2000, Fabian et al. 2001). The
curvature can be accounted for by excess absorption below 1~keV over
the Galactic value, and indeed, medium-hard X-ray observations of
high-redshift radio-loud quasars has confirmed the presence of
spectral flattening at the lower energies (Page et al. 2005; Reeves \&
Turner 2000; Cappi et al. 1997). Finally, what makes this energy range
of interest is the possibility to detect signatures of a
matter-dominated jet (Sikora et al. 1997). The latter is predicted to
show a downward-curved continuum in 0.3--10~keV due to the presence of
a thermal (blackbody) component related to bulk Compton scattering of
the Broad Line Regions photons off the cold particles in the jet
(Celotti, Ghisellini, \& Fabian 2006).

The advent of sensitive hard X-ray experiments, such as the Burst
Alert Telescope (BAT; Barthelmy et al. 2005) in 15--150~keV onboard
\swift\ (Gehrels et al. 2004) is changing our understanding of AGN
jets. What makes the BAT of interest is its large field of view and
survey mode operation, which allows detection of a large number of
relatively bright sources, such as blazars. In fact, during its
3-month survey the BAT (Markwardt et al. 2005) discovered a new, 
MeV-bright blazar at
$z$=2.979, SWIFT~J0746.3+2548 (Sambruna et al. 2006a, S06 in the
following).  Motivated by this discovery, we searched the BAT AGN
9-months archive (Tueller et al. 2007, in prep.) for additional $z>2$
radio-loud quasars, and found four: 0212+735, 0537--286, 0836+710, and
2149--307. From previous observations these quasars are known X-ray
emitters at soft-medium X-ray energies (see references in Table~5).

Here we report the analysis and interpretation of the BAT spectra, and
follow-up XRT (Burrows et al. 2005) and UVOT (Roming et al. 2005)
observations of the four sources.  After the description of the
\swift\ observations and data analysis in
\S~2, and timing and spectral analysis in \S~3, we discuss the SEDs
and their modeling in \S~4. Summary and conclusions follow in \S~5. We
assume a concordance cosmology with H$_0=71$ km s$^{-1}$ Mpc$^{-1}$,
$\Omega_{\Lambda}$=0.73, and $\Omega_m$=0.27 (Spergel et
al. 2003). The photon index $\Gamma$ is defined such that
$\Gamma=\alpha+1$, where $\alpha$ is the energy index, $F_{\nu}
\propto \nu^{-\alpha}$.

\section{Observations} 

\subsection{The sources} 

The quasars analyzed here are included in the list of identified
extragalactic sources detected by BAT up to August 2006. They are
0212+735, 0537--286, 0836+710, and 2149--307, known blazars at
redshifts $z>2$. With the exception of 0212+735, all sources are
well-known emitters at medium-hard X-rays (Donato, Sambruna, \&
Gliozzi 2004), with hard continua, photon indices
$\Gamma=1.3-1.5$. Two of them, 0836+710 and 2149--307, were detected
with the \sax\ PDS up to 100~keV (Tavecchio et al. 2000, Elvis et
al. 2003); 0836+710 was also detected at GeV gamma-rays with EGRET
onboard CGRO (Hartman et al. 2003) and with \integral\ (Beckmann et
al. 2006).

Table~1 lists the sources with their basic properties. Also listed are
the BAT count rates in 15--150~keV from the 9-month survey, and the
count rates from the follow-up XRT observations in 0.3--10~keV. All the
four sources were observed more than once with the XRT, with little or
no variability. 

The sources were also observed at various epochs with the UVOT. As no
flux variability is present for each source at the various epochs, the
magnitudes were coadded. They are listed in Table~2, together with the
flux densities. The magnitudes were corrected for Galactic extinction
(see below). 

\subsection{The X-ray data} 

\noindent{\bf BAT:}
The reduction of the BAT data was described in S06. Table~1 shows the
BAT count rates; the sources are detected at $>5\sigma$. Spectra were
integrated for the first 9 months of the BAT survey, and fitted with
\verb+XSPEC+ v.11.3.2, with the response matrix based on the latest
(August 2006) calibration products. Count rates were collected in four
energy bands (14--25~keV, 25--50~keV, 50--100~keV, and 100--200~keV),
and converted into fluxes using a standard diagonal response.

\noindent{\bf XRT:} The unfiltered data were downloaded from the public archive
and processed using \verb+xrtpipeline+ v0.10.6. Standard grade
selection, 0--12 for Photon Counting mode (Hill et al. 2004), was used
for both spectral and timing analysis. Light curves and spectra were
extracted from a circular region centered on the XRT position and with
radius (45--55)\arcsec. The background was extracted from an annulus
centered on the source and with inner and outer radii (65--80)\arcsec\
and (125--135)\arcsec, respectively. Inspection of the background
light curve shows no variability. Light curves and spectra were
corrected for the exposure map, accounting for CCD bad columns and
pixels, hot pixels, attitude variations, and telescope vignetting.

The XRT spectra were fitted within \verb+XSPEC+ v.11.3.2ad in the
energy range 0.3--10~keV.  We used the latest spectral redistribution
matrices (RMF, v008). Ancillary response files were generated with the
standard \verb+xrtmkarf+ tool v0.5.2 using calibration files from
\verb+CALDB+ v2.6 (as part of the \verb+HEAsoft+ v6.1.2), and corrected for
the exposure map.  No pileup is present at the sources' count
rates. The XRT spectra were rebinned in order to have at least 20
counts in the new bins, to enable use of the $\chi^2$ statistics.
After checking the consistency of the data at the various epochs, we
performed joint fits to the XRT spectra for each source to improve the
signal-to-noise ratios.

\subsection{The optical/UV photometry} 

The data analysis was performed using the \verb+uvotsource+ task
included in the software release \verb+HEAsoft+ v6.0.5. Photometry was
done using an aperture with a radius of 2\arcsec\ and aperture
corrected to the standard UVOT photometry apertures for point sources
(6\arcsec\ for the UBV filters and 12\arcsec\ for the UV filters).
The magnitudes were corrected for absorption and converted into the
fluxes in Table~2. The adopted reddening corrections are E(B-V)=0.744,
0.025, 0.101, and 0.025 mag for 0212+735, 0537-236, 0836+715, and
2149-307, respectively (Schlegel et al. 1998). While 0836+710 and
2149--307 are well detected with all the optical and UV filters,
0212+735 and 0537--286 were not detected in the UV, where 3$\sigma$
upper limits are given for the flux densities.

\section{X-ray Analysis} 

\subsection{Timing Analysis} 

Within the short XRT exposures there is no evidence for significant
flux variability. The only exception is the second observation of
2149--307, where a flux change of a factor 2 in $\sim$ 3 hours is
observed. The variability is significant at 99.2\% confidence
according to the $\chi^2$ test. Comparison of the 2--10~keV flux with
previous \asca, \sax, and \xmm\ observations reveals that the XRT
observed the sources in a typical average state (\S~3.2.4).

Light curves were also extracted in the energy range 15--150~keV, and
binned with 28-day bins. No episodes of outburst are present in the
BAT 9-months light curves; a $\chi^2$ test returns values of
$\chi^2$=1.23, 3.34, 3.1, and 2.99 for 0212+735, 0537--286, 0836+710,
and 2149--307, respectively. Because of the large statistical errors,
the minimum $\chi^2$ value indicating possible variability is
$\chi^2=3$. Based on this criterion, evidence for flux variations is
present only in 0836+710 (Fig. 1), where the 15--150~keV flux
increased by a factor 4 on a timescale of one month, followed by a
similar-amplitude decay in 3--4 months.

\subsection{Spectral Fits}

Table~3 reports the results of the spectral fits to the BAT and XRT
spectra of the four sources. The best-fit parameters and their 90\%
uncertainties for one parameter of interest ($\Delta\chi^2$=2.7) are
listed, together with the observed fluxes and intrinsic
(absorption-corrected) luminosities. The significance of adding free
parameters to the model was evaluated with the F-test, with related
probability P$_F$.

\subsubsection{BAT} 

The BAT spectra were fitted with a single power law with no
absorption, as the latter, for the small column densities involved in
this work, has negligible effects above 10~keV. The results of the
fits are listed in Table~3a. The fits with the power law model are
satisfactory, yielding slopes in the range $\Gamma_{BAT}
\sim 1.3-1.8$. 

\subsubsection{XRT} 

The XRT spectra of the sources were fitted in 0.3--10~keV at first
with a single power law model plus a column density N$_H$ fixed to the
Galactic value. Inspection of the residuals of this model (Figure~2)
reveals spectral flattening below 2~keV in all sources. This indicates
a curved continuum, either because of excess absorption over Galactic
at the lower energies, or because of intrinsic curvature.

We thus fitted the XRT continua with two phenomenological models: 1) a
power law with an extra absorber, N$_H^z$, assumed to be at the
source's rest-frame; and 2) a broken power law. The Galactic column
density was included in both models. The results for both models are
listed in Table~3b. Generally, models 1 and 2 both provide an equally
acceptable fit except in the case of 0836+710, where the curved broken
power law is preferred over the absorbed power law at P$_F$=99\%. 

\subsubsection{Joint fits to XRT+BAT} 

Joint spectral fits to the XRT and BAT data were also attempted. Even
before formally fitting, it was apparent that the BAT data lied on the
extrapolation of the XRT best-fit model, both for the absorbed power
law and the broken powerlaw. An exception is 0836+710, where the BAT
lie above the extrapolation of the XRT power law, as expected. 

The 0.3--150~keV spectra were formally fitted with the absorbed power
law model (broken powerlaw for 0836+710), which gave fitted parameters
very similar to those of the XRT continuum alone; Table~3c reports
these fits, while the data are shown in Figure~3. Similarly, the
broken power law fits (not reported) gave similar parameters to the
XRT alone. We also searched for the Compton bump by adding a blackbody
to the power law, with no fit improvement. For 0836+710, however, the
addition of a blackbody with fitted temperature $kT \sim 3$ keV
provides much better BAT residuals. More sensitive observations, such
as those already performed with \suzaku, are needed to confirm the
possible Compton bump in this object.

\subsubsection{Comparison to previous X-ray observations} 

Table~4 presents a compilation of previous observations at X-rays of
the sources, specifically, X-ray fluxes and photon indices measured by
previous missions in the energy bands 2--10~keV (\asca,
\xmm, \sax) and $>$ 10~keV (\sax, \integral). Only 0537--286, 0836+710, and
2149--307 were observed above 2 keV; the only previous observation of
0212+735 was during the \rosat\ All-Sky Survey. During the RASS,
0212+735 had a count rate of 0.0437~c/s in 0.2--2.4~keV, consistent
with the extrapolation of the XRT best-fit model to these energies.

Inspection of Table~4 reveals modest-amplitude (factor 2) flux
variability at medium-hard X-rays, with little or no spectral
changes. The only hint for correlated flux and spectral changes is for
0836+710: between the \sax\ and \swift\ observations the flux
decreased slightly (factor 1.6) with the continuum becoming softer.

Above 10~keV, only 0836+710 and 2149--307 were previously observed with \sax\
(Tavecchio et al. 2000; Elvis et al. 2003). In 0836+710, the 20--200~keV
flux decreased by a factor 5 between the PDS and BAT epochs while the
continuum steepened from $\Gamma \sim 1.4$ to $\Gamma \sim 1.8$. Only
modest-amplitude flux variations, factor 1.5, are observed for 2149--307,
with no spectral changes.

\section{Discussion} 

\subsection{X-ray spectra of high-redshift radio-loud quasars} 
 
X-ray observations of high-redshift quasars with \rosat\ and \asca\
found evidence for significant excess column densities in radio-loud
but not in radio-quiet sources (Cappi et al. 1997, Reeves
\& Turner 2000). It was claimed that excess absorption is a
common property of radio-loud quasars only, and as such most likely of
intrinsic origin. Based on multi-epoch X-ray spectroscopy, it was also
claimed that absorber is variable.

Indeed in several high-redshift blazars ($z>4$) the steepening of the
soft X-ray spectrum is interpreted (e.g., Worsley et al. 2006 and
references therein) in terms of absorption by a dense region of warm
plasma ($N_H\sim 10^{22}$ \nh) present in the region surrounding the
QSO, and expected to be in the form of a wind/outflow (e.g., Fabian
1999). However, the absorption scenario poses several difficulties. In
particular, it appears difficult to reconcile the presence of a large
gas column on the line of sight to the jet, which is expected to
efficiently remove gas from its vicinity, in particular in the case of
blazars for which the line of sight must be close to the axis of the
jet. 

As mentioned before, a viable alternative to absorption is to assume
that continuum is {\it intrinsically curved}. Indeed, a hardening of
the continuum is naturally expected in the soft X-ray band, where the
EC emission is produced by the relativistic electrons with the lowest
energy. The exact shape of the spectrum in this region depends on the
detailed energy distribution of the electrons and on the spectrum of
the soft photons, but the overall shape could easily mimick the effect
of absorption (for more details see Tavecchio et al. 2007). The
hardening of the EC component in the soft band can be hidden by the
presence of the SSC component. From the observational point of view,
however, the contribution of the SSC decreases for increasing jet
power; in the most powerful sources, such as discussed in this paper,
the EC component is basically ``naked'' and the break can be observed.

Celotti et al. (2006) discussed the possible the presence of a weak
thermal bump around 10~keV produced by inverse Compton scattering of
external photons by {\it cold} leptons in the jet approaching the BLR
region (the so-called bulk Compton emission). 
The direct detection of such a component, however, has remained
elusive, since the emission is dominated by the strong non-thermal
emission (SSC, EC). While \suzaku\ is more sensitive at the higher
energies than the BAT, the latter has the advantage of a continuous
scan of the sky with a large field of view. This allows for the study
of a larger number of sources, and - on the long run - the possibility
to detect faint ones as well. However, clearly \suzaku\ remains the
instrument of choice for a quasi-instantaneous snapshot of the
broad-band X-ray spectrum.

\subsection{Spectral Energy Distributions} 

The Spectral Energy Distributions (SEDs) of the four sources,
assembled from Tables 1 and 2 and from observations at longer
wavelengths from the literature, are shown in Figure~4. In all the
cases we report the X-ray data obtained assuming that the continuum is
intrinsically curved (i.e., without extra-absorption; see Tavecchio et
al. 2007). For 0836+710, the only source detected at GeV gamma-rays
with EGRET (Hartman et al. 2003), the GeV data are shown as well while
for the remaining sources 3$\sigma$ GeV upper limits are shown, based
on the EGRET sensitivity. Inspection of Figure~4 reveals the typical
double-peaked SEDs of blazars (\S~1) with the first hump due to
synchrotron emission from relativistic electrons in the jet, and the
second component due to IC, either SSC or/and EC.

Our four sources are technically classified as Flat Spectrum Radio
Quasars (FSRQs), as they are known to exhibit strong broad emission
lines in their optical spectra. The SEDs in Figure~4 exhibit a very
steep optical-to-UV continuum. This strongly suggests that the
emission in this band is dominated by a thermal-like component,
possibly associated with the blue-bump component that characterizes
the quasar optical region (e.g., Sun \& Malkan 1989), usually thought
to be associated with the putative accretion disk.  If we assume that,
as in most FSRQs, the peak of the synchrotron emission is at IR, than
we are forced to conclude that the optical-UV emission in the SEDs in
Figure~4 has a different origin than the longer wavelengths. This is
similar to the case of SWIFT~J0746.3+2548 (S06), where the emission in
optical-UV was also interpreted as due to the thermal contribution of
the accretion disk/BLRs.

Note that in 0212+735 and in 0537--286 the optical-to-UV peak is very
narrowly defined, due to a sudden increase of the flux in the B filter
(Table~2). A possibile explanation for this increase could be that the
continuum flux in the B-band is strongly contaminated by the
redshifted Ly$\alpha$ emission line, which is usually strong in
luminous quasars (Lanzetta, Turnshek, \& Sandoval 1993). It is also
possible that the steep cutoff in the UV could be due, at least in
part, to intervening absorption; for 0537--286, a strong continuum
discontinuity attributed to the the Lyman continuum edge was detected
at 3650~\AA\ in an early optical spectrum, implying an intervening
column density $\sim 10^{17}$ \nh\ (Wright et al. 1978). Ideally,
optical-UV polarimetry could ascertain the origin of the optical-UV
continuum in these sources, distinguishing between thermal (disk) and
non-thermal, highly polarized (jet) emission. Such observations in
case of 0836+710 yielded a polarimetry degree of $1.1 \pm 0.5$\%
(Impey \& Tapia 1990), lower than usually observed in the
jet-dominated continuum of other blazars.

The hard BAT spectra, together with the limits from EGRET at GeV
energies, imply a peak of the IC component around a few GeV. This is
in contrast to SWIFT~J0746.3+2548, detected during a flaring state
with BAT (S06). In this source, the exceptionally hard BAT spectrum
($\Gamma_{BAT} \sim 1$) and the previous EGRET non-detection imply a
peak of the IC component around MeV energies. Thus, it could be the
case that during hard X-ray outbursts the IC peak moves from GeV to
MeV energies, but only detailed monitoring at the higher energies can
address this issue. This will soon be possible with \glast.

\subsection{Modeling the SEDs} 

We reproduce the multiwavelength SEDs of the four quasars in Figure~4
using a simple, one-zone, homogeneous synchrotron and inverse Compton
model, fully described in Ghisellini, Celotti, \& Costamante
(2002). The basic assumptions of the model are:

\noindent
1) The source, moving with a bulk Lorentz factor $\Gamma$ and seen at
an angle $\theta$, is a cylinder of cross sectional radius $R=\psi z$,
where $z$ is the distance from the apex of the jet, assumed conical of
semiaperture angle $\psi$. The width $\Delta R^\prime$, as measured in
the comoving frame, is here assumed to be equal to $R/\Gamma$. The
magnetic field $B$ is homogeneous and tangled.
 
\noindent
2) The external radiation is produced at a fixed radius, identified
with the radius of the broad line region, $R_{\rm BLR}$.  We choose to
model this component with a black body with luminosity $L_d$ peaking
at the typical frequency of $10^{15}$ Hz (in the quasars'
rest-frame). This component approximates the radiation reprocessed and
re-isotropized by the BLR clouds. We assume that 10\% of the disk
luminosity is reprocessed by the broad line region. Note that, besides
the contribution of the BLR, other processes can contribute to the
external radiation, such as scattering by ionized intercloud plasma,
synchrotron radiation ``mirrored" by the clouds and/or the walls of
the jet, reprocessing by a molecular torus (Ghisellini \& Madau
1996). Inclusion of these second-order effects is not warranted by the
limited data coverage, especially at IR wavelengths.

\noindent
3) Relativistic electrons are injected with luminosity
$L^{\prime}_{\rm inj}$ and the distribution is the result of injection
and cooling.  We calculate for which $\gamma=\gamma_{\rm cool}$ the
particles cool in one light crossing time.  If the particles are
injected between $\gamma_1$ and $\gamma_2$ with a power law
distribution of slope $s$, in the ``fast cooling" regime ($\gamma_{\rm
  cool}<\gamma_1$), we have $N(\gamma) \propto \gamma^{-p}$ between
$\gamma_1$ and $\gamma_2$, (where $p=s+1$ is the injection slope
increased by one unit) and $N(\gamma) \propto \gamma^{-2}$ between
$\gamma_{\rm cool}$ and $\gamma_1$.  In the models presented here,
this is always the case, since $\gamma_{\rm cool}$ is of order unity
in all cases.

The results of the fits to the SEDs with this model are shown in
Figure~4 (solid lines). The assumed parameters are reported in 
Table~5. In all cases we interpret the UVOT data as tracing the
high-frequency tail of the disk emission; this allows us to fix the disk
luminosity $L_d$. For all the sources we assume that the peak of the
IC component is just above the BAT band: with this choice
(consistent with the non-detection in the EGRET band for three of the
objects) we minimize the emitted radiative power of the jets. Note
that large variations in the $\gamma$-ray band could be easily
obtained by changing the upper end of the electron energy distribution
$\gamma_2$, without appreciable variations in the X-ray band.

Assuming the composition of one proton per emitting electron (for the
reason of this choice see Maraschi \& Tavecchio 2003), we infer jet
kinetic powers of $P_j \sim 10^{48}-10^{49}$ \lum. This can be
compared with the radiative output of the disk $L_d \sim 10^{47}$
\lum. The ratio $L_d/P_j \sim 0.01-0.1$ is of the same order of
magnitude as found in other cases for which both powers can be
estimated with reasonable accuracy (Tavecchio et al. 2000; Maraschi \&
Tavecchio 2003). The parameters used to reproduce the SEDs are close
to those usually found for other powerful blazars. 

Figure~4 shows that in the four quasars the jet component dominates
the high-energy emission above a few hundred eV. It is interesting to
compare with intermediate-redshift radio-loud quasars. Analysis of the
core SEDs of three blazars at z=0.3--0.8 with detected \chandra\ jets
(Sambruna et al. 2006b) shows that in these sources, the jet
contributes only 50\% of the X-ray emission in 0.3--10~keV, becoming
dominant at higher energies. A narrow \feka\ line at 6~keV with EW
$\sim$ 100 eV was in fact detected in the three sources, as well as an
optical-UV bump related to the accretion disk. Thus, it appears that
the high-energy spectra of radio-loud quasars may contain significant
contribution from thermal emission of the accretion disk; however, the
jet becomes increasingly dominant at X-ray energies with increasing
jet power/$z$. This is not unexpected if selection biases due to
beaming are at work: only the most beamed, and thus jet-dominated,
radio-loud quasars can be detected at cosmological distances.

The sources discussed in this work are quite similar to RBS 315,
another high-redshift ($z=2.69$) FSRQ recently observed by us with
\suzaku\ (Tavecchio et al. 2007). In this case as well the X-ray spectrum
flattens in the soft X-ray band, requiring either absorption or
intrinsic curvature of the underlying continuum. A previous \xmm\
observation of RBS~315 revealed an extremely hard X-ray continuum,
with photon index $\sim 1.2$, requiring an extremely flat electron
distribution. Very hard spectra ($\Gamma <1.5$) are also displayed by
our sources (Table~4).

\section{Summary and Conclusions} 

Using the BAT experiment on-board \swift, we detected hard ($>$
10~keV) emission from four $z>2$ radio-loud quasars. From the modeling
of the (non-contemporaneous) SEDs, we find that the jet dominates the
emission at higher energies, while the steep optical-UV continuum is
readily interpreted as due to the accretion disk. Polarimetry of the
optical and UV continuum from these quasars is encouraged to confirm
this intepretation. The disk luminosity is 1\%--10\% the power in the
jet.

High-redshift blazars qualify as excellent candidates to investigate
the disk-jet connection in radio-loud AGN. Soon another very important
piece of the puzzle will be available, thanks to the launch of the
\glast\ observatory. The synergy between \glast\ and \swift\ 
holds the key to elucidate the origin of the high-energy radiation
from powerful radio-loud quasars.

\acknowledgements

This research has made use of data obtained from the High Energy
Astrophysics Science Archive Research Center (HEASARC), provided by
NASA's Goddard Space Flight Center, and of the NASA/IPAC Extragalactic
Database (NED) which is operated by the Jet Propulsion Laboratory,
California Institute of Technology, under contract with the National
Aeronautics and Space Administration.


\clearpage 


\scriptsize
\begin{center}
\begin{tabular}{lllllllll}
\multicolumn{9}{l}{{\bf Table 1: $z>2$ radio-loud quasars detected by BAT}} \\
\multicolumn{9}{l}{   } \\ \hline
& & & & & & & & \\
Source & RA(2000) &  DEC(2000)  &$z$ & N$_H$ & BAT & Date & Exp & XRT  \\ 
& & & & & & & & \\
(1) & (2) & (3) & (4) & (5) & (6) & (7) & (8) & (9) \\
& & & & & & & & \\ \hline
& & & & & & & & \\
0212+735 (1) &  02 17 30.8 & +73 49 33 &  2.367 & 27.0 & 3.8 $\pm$ 0.9 & 2005 December 1 & 4.6 & 6.6 $\pm$ 0.4 \\
~~~~~~~~~~~~ (2) &   & &     &      &       & 2006 July 18 & 6.5 & 6.9 $\pm$ 0.3 \\
~~~~~~~~~~~~ (3) &   & &     &      &       & 2006 July 19 & 1.0 & 6.8 $\pm$ 0.8 \\

0537--286 (1) & 05 39 54.3 & --28 39 56  & 3.104 & 1.9 & 4.4 $\pm$ 0.9 
                            & 2005 November 24 & 9.2 & 5.0 $\pm$ 0.2 \\
~~~~~~~~~~~~~~(2) & & & & & & 2005 December 8 & 14.7 & 5.5 $\pm$ 0.2 \\
~~~~~~~~~~~~~~(3) & & & & & & 2006 October 27 & 3.0 & 5.3 $\pm$ 0.5 \\
~~~~~~~~~~~~~~(4) & & & & & & 2006 October 30 & 3.9 & 6.7 $\pm$ 0.4 \\
~~~~~~~~~~~~~~(5) & & & & & & 2006 October 31 & 3.4 & 5.0 $\pm$ 0.2 \\
~~~~~~~~~~~~~~(6) & & & & & & 2007 May 17 & 5.4 & 6.7 $\pm$ 0.3 \\

0836+710 (1) & 08 41 24.3 & +70 53 42 & 2.172 & 2.9 & 10.2 $\pm$ 0.9  & 2006 April 1 & 4.4 & 37.9 $\pm$ 0.9 \\ 
~~~~~~~~~~~~~(2) & & & & & & 2006 July 10 & 1.9 & 28.1 $\pm$ 1.0 \\
~~~~~~~~~~~~~(3) & & & & & & 2007 April 4 & 7.3 & 28.9 $\pm$ 0.6 \\

2149--307 (1) & 21 51 55.5 & --30 27 54 & 2.345 & 1.9 & 9.5 $\pm$ 1.6 & 2005 December 10 & 3.1 & 31.2 $\pm$ 1.0 \\
~~~~~~~~~~~~ (2) &       & & &     &                  & 2005 December 13 & 2.3 & 25.1 $\pm$ 1.1 \\
& & & & & & & & \\ \hline

\end{tabular}
\end{center}

\noindent
{\bf Explanation of Columns:} 
1=Source name; 
2=RA(2000) in hh mm ss;  
3=DEC(2000) in dd mm ss; 
4=Redshift; 
5=Galactic column density (in $10^{20}$ \nh);
6=Count rate in 15--150~keV from the 9-months BAT survey in $10^{-5}$ counts s$^{-1}$;
7=Date of the XRT follow-up observation;
8=XRT net exposure after data screening (in ks); 
9=Count rate in 0.3--10~keV in 10$^{-2}$ counts s$^{-1}$. 

\normalsize


\clearpage

\scriptsize
\begin{center}
\begin{tabular}{lllllr}
\multicolumn{6}{l}{{\bf Table 2: UVOT Observations}} \\
\multicolumn{6}{l}{   } \\ \hline
Source & Filter & Wavelength & Mag & Flux & Exposure \\
& & & & & \\
(1) & (2) & (3) & (4) & (5) & (6)  \\ \hline
&& & &  & \\ 
0212+735 &V & 5402 & 17.60 $\pm$ 0.26 & 2.9 $\pm$ 0.7 & 1006 \\
         &B & 4331 & 16.49 $\pm$ 0.14 & 10.2 $\pm$ 1.3 & 924 \\
         &U & 3450 & 16.86 $\pm$ 0.25 & 6.0 $\pm$ 1.4 & 1004 \\
      &UVW1 & 2634 & $>$ 16.85 & $<$ 17.7  & 2344 \\
      &UVM2 & 2231 & $>$ 15.04 & $<$ 91.0 & 2789 \\
      &UVW2 & 2034 & $>$ 16.22 & $<$ 34.6  & 4088 \\ 
& & & & & \\
0537--286 & V & 5402 & 18.99 $\pm$ 0.07 & 0.80 $\pm$ 0.05 & 1995 \\
     & B & 4331 & 19.58 $\pm$ 0.06 & 0.89 $\pm$ 0.05 & 1436 \\
     & U & 3507 & 20.59 $\pm$ 0.14 & 0.19 $\pm$ 0.03 & 1962 \\
       &UVW1 & 2634 & $>$ 22.18 & $<$ 0.06 & 3984 \\
       &UVM2 & 2231 & $>$ 22.52 & $<$ 0.06 & 6086 \\
       &UVW2 & 2034 & $>$ 22.74 & $<$ 0.06 & 8108 \\ 
& & & & & \\
0836+710 & V & 5402 & 16.49 $\pm$ 0.07 & 8.1 $\pm$ 0.52 &  372 \\
         & B & 4331 & 16.66 $\pm$ 0.06 & 13.1 $\pm$ 0.7 & 372 \\
         & U & 3507 & 16.31 $\pm$ 0.07 & 12.5 $\pm$ 0.8 & 372 \\
         & UVW1& 2634 & 16.31 $\pm$ 0.07 & 12.5 $\pm$ 0.8 & 746 \\
         & UVM2& 2231 & 17.15 $\pm$ 0.10 & 7.8 $\pm$ 0.7 & 948 \\
         & UVW2& 2034 & 16.96 $\pm$ 0.05 & 12.7 $\pm$ 0.6 & 1518 \\
& & & & & \\ 
2149--307  & V & 5402 & 17.31 $\pm$ 0.06 & 3.8 $\pm$ 0.2 & 441 \\
     & B & 4331 & 17.52 $\pm$ 0.05 & 6.0 $\pm$ 0.3 & 287 \\
     & U & 3507 & 16.98 $\pm$ 0.05 & 5.4 $\pm$ 0.3 & 441 \\
     & UVW1& 2634 & 17.98 $\pm$ 0.07 & 2.7 $\pm$ 0.2 & 879 \\
     & UVM2& 2231 & 20.40 $\pm$ 0.22 & 0.4 $\pm$ 0.08 & 1322 \\
     & UVW2& 2034 & 19.99 $\pm$ 0.13 & 0.8 $\pm$ 0.1 & 2120 \\ 
& & & & & \\ \hline

\end{tabular}
\end{center}

\noindent
{\bf Explanation of Columns:} 
1=Source name; 
2=Filter; 
3=Center wavelength in \AA; 
4=Observed magnitude; 
5=Flux density in $10^{-16}$ erg cm$^{-2}$ s$^{-1}$ \AA$^{-1}$, dereddened for Galactic absorption. Upper limits are 3$\sigma$; 
6=Exposure in s.

\normalsize

\clearpage
\def\flux{erg cm$^{-2}$ s$^{-1}$}


\scriptsize
\begin{center}
\begin{tabular}{lllllllllll}
\multicolumn{11}{l}{{\bf Table 3: Spectral fits}} \\
\multicolumn{11}{l}{   } \\ \hline
& & & & & & & & & & \\
& \multicolumn{3}{c}{Power Law} &&  \multicolumn{4}{c}{Broken Power Law} & Flux & Lum\\
& & & & & & & & & & \\ \hline
Source & N$^z_H$ & $\Gamma$ & $\chi^2_r$/dofs && $\Gamma_1$ & $\Gamma_2$ & E$_b$& $\chi^2$/dofs & & \\
& & & & & & & & & & \\
(1) & (2) & (3) & (4) && (5) & (6) & (7) & (8) & (9) & (10) \\
& & & & & & & & & & \\ \hline
& & & & & & & & & & \\
\multicolumn{11}{l}{{\bf 3a: Fits to the BAT continuum in 15--150~keV}} \\ \hline
0212+735 & $\cdots$  & 1.84$^{+0.57}_{-0.68}$ & 0.63/1 && & & & & 23 & 78 \\
0537--286 & $\cdots$ & 1.26$^{+0.35}_{-0.75}$ & 0.97/2 && & & & & 49 & 86 \\
0836+710 & $\cdots$ & 1.83$^{+0.24}_{-0.23}$ &0.23/2  & & & & & &65.9 & 95.7 \\
2149--307 & $\cdots$ & 1.44$^{+0.43}_{-0.42}$ & 0.58/2 && & & & & 88.0 & 110 \\
& & & & & & & & & & \\ \hline
& & & & & & & & & & \\

\multicolumn{11}{l}{{\bf 3b: Fits to the XRT continuum in 0.3--10~keV}} \\ \hline

0212+735 & 6.5$^{+4.8}_{-3.9}$ & 1.32$^{+0.19}_{-0.17}$ & 1.22/41 && 0.9$^{+0.17}_{-3.9}$ &
         3.9$^{+2.2}_{-3.7}$ & 1.7$^{+1.2}_{-0.6}$ & 1.25/40 & 5.6--6.9 & 7.5--14.0 \\

0537--286 & 1.3$^{+0.7}_{-0.6}$ & 1.36 $\pm$ 0.07 & 1.09/105 && 0.43$^{+0.53}_{-0.56}$ &
             0.9$^{+0.6}_{-0.1}$ & 1.35 $\pm$ 0.05 & 0.90/102 &
                  2.9--3.2 & 10.1--11.3 \\

0836+710 & 0.81$^{+0.44}_{-0.41}$ & 1.53 $\pm 0.06$ & 1.09/168 && 1.15$^{+0.11}_{-0.12}$ &
           1.7$^{+0.5}_{-0.3}$ & 1.61$^{+0.12}_{-0.09}$  & 1.05/167 & 16.7--20.7 & 29.6--36.8 \\

2149--307  & 0.25$^{+ 0.34}_{-0.25}$  & 1.50 $\pm 0.09$ & 0.97/66 && 1.25 $\pm$ 0.20 &
       1.3$^{+1.8}_{-0.4}$ & 1.56$^{+0.31}_{-0.10}$ & 0.95/65 & 11.9--14.8 & 25.2--31.3 \\ 
& & & & & & & & & & \\\hline 
& & & & & & & & & & \\

\multicolumn{11}{l}{{\bf 3c: Fits to the XRT + BAT continuum in 0.3--150~keV}} \\ \hline

0212+735 & 8.8$^{+3.8}_{-3.2}$ & 1.43$^{+0.11}_{-0.09}$ & 1.20/44 && & & & & & \\

0537--286 & 1.4 $\pm$ 0.6 & 1.38 $\pm$ 0.07 & 0.90/106 && & & & & & \\

0836+710 &&&&& 1.16 $\pm$ 0.13 & 1.70$^{+0.22}_{-0.13}$ & 1.8$^{+0.4}_{-0.2}$ & 1.04/168 && \\

2149--307 & 0.25$^{+0.34}_{-0.25}$  & 1.50 $\pm 0.09$ & 0.94/69 && & & & & & \\
& & & & & & & & & & \\\hline

\end{tabular}
\end{center}

\noindent
{\bf Explanation of Columns:} 
1=Source name; 
2=Column density at the source's rest-frame ($10^{22}$ \nh); 
3=Photon Index; 
4=Reduced $\chi^2$ and degrees of freedom; 
5=Photon Index below the break;
6=Photon Index above the break;
7=Break energy (keV);
8=Reduced $\chi^2$ and degrees of freedom;
9=Observed flux in 20--200~keV (BAT) and
  2--10~keV (XRT) from the best-fit model (10$^{-12}$ \flux);
10=Intrinsic (absorption-corrected) luminosity in 20--200~keV (BAT) and
  2--10~keV (XRT) from the best-fit model ($10^{46}$ \lum).

{\bf Notes:} a=All components are absorbed by the Galactic column density. 
 
\normalsize

\clearpage

\scriptsize
\begin{center}
\begin{tabular}{lllllllllcc}
\multicolumn{11}{l}{{\bf Table 4: Swift and Previous X-ray Observations}} \\
\multicolumn{11}{l}{   } \\ \hline
& & & & & & & & & & \\ \hline 
& & & & & & & & && \\
\multicolumn{11}{l}{{\bf a: In the 2--10~keV energy range}} \\ \hline 
& & & & & & & & && \\
Source && \asca\ && \sax\ && \xmm\ && \swift\ && Refs. \\ 
& & & & & & & & && \\ \hline
& & & & & & & & && \\
0537--286 & Flux$^a$ & 2.04 &&  $\cdots$ && 2.93 && 2.9--3.2 && 1,2 \\
 & $\Gamma$  & $1.47 \pm 0.08$ && $\cdots$ && 1.41 $\pm$ 0.01 && 
                 1.36 $\pm$ 0.07  && \\

0836+710 & Flux$^a$ & 18.5 && 26.0 && 21.52 && 16.7--20.7 && 1,3,2 \\
 & $\Gamma$ & $1.41 \pm 0.03$ && 1.33 $\pm$ 0.04 && 1.37 $\pm$ 0.01 && 
                 1.53 $\pm 0.06$ && \\

2149--307 & Flux$^a$ & 14.05 && 8.0 && 3.5 && 11.9--14.8 && 1,4,2 \\
 & $\Gamma$ & $1.49 \pm 0.04$ && 1.4 $\pm$ 0.04 && 1.47 $\pm$ 0.01 &&  1.50 $\pm 0.09$ \\
& & & & & & & & && \\ 
\multicolumn{11}{l}{{\bf b: In the 20--200~keV energy range}} \\ \hline 
& & & & & & & & && \\
Source &&  && \sax\ && \integral\ && \swift\ && Refs. \\ 
& & & & & & & & && \\ \hline
& & & & & & & & && \\
0836+710 & Flux$^a$ &   &&  230 && 77.7 && 65.9 && 3,5 \\
          & $\Gamma$ &   &&  1.43 $\pm$ 0.05 && 1.65$^{+0.46}_{-0.39}$ && 
                       1.83$^{+0.24}_{-0.23}$ && \\

2149--307 & Flux$^a$ &  && 57.0 &&  && 88.0 && 4 \\ 
          & $\Gamma$ &   && 1.40 $\pm$ 0.04 && && 1.44$^{+0.43}_{-0.44}$ && \\ \hline
& & & & & & & & && \\

\end{tabular}
\end{center}

\noindent
{\bf Notes:} 
a=Observed flux in $10^{-12}$ \flux. 

{\bf References:} 1=Reeves \& Turner (2000); 2=Page et al. (2005); 3=Tavecchio et al. (2000);
4=Elvis et al. (2000); 5=Beckmann et al. (2006).

\normalsize

\clearpage


\begin{center}
\begin{tabular}{rlrlllllllr}
\multicolumn{11}{l}{{\bf Table 5: SEDs Parameters}} \\
\multicolumn{11}{l}{ } \\ 
\hline 
 & & & & & & & & & & \\ 
Source & $R$ & $\Gamma$ & $\theta$ & $B$ & $L^{\prime}_{\rm inj}$ & $\gamma_1$ & $\gamma_2$ & $s$ & $L_d$ & $R_{BLR}$ \\ 
 & & & & & & & & & &\\ 
(1) & (2) & (3) & (4) & (5) & (6) & (7) & (8) & (9) & (10) & (11) \\ 
 & & & & & & & & & &\\ 
\hline 
 & & & & & & & & & &\\

0212+735 & 1.5& 12& 3.0& 10&8& 100& 2$\times 10^3$& 2.5& 4  & 12\\

0537-286 & 2.0& 15& 3.0& 2.5&13& 20& $10^4$        & 2.9& 0.6& 4.5\\

0836+710 & 1.5& 13& 3.3& 12&12& 6 & 5$\times 10^3$& 3.0& 2.8& 15\\

2149-307 & 1.0& 17& 3.0& 7 &6 & 4.5& $10^3$       & 2.9& 1.5& 8.0\\ 
 & & & & & & & & & \\ \hline

\end{tabular}
\end{center}

\noindent 

{\bf Explanation of Columns:} 1=Source name; 
2=Radius $R$ of emitting region in units of $10^{16}$ cm;
3=Bulk Lorentz factor;
4=Viewing angle (degrees);
5=Magnetic field intensity (G);
6=Intrinsic injected power in units of $10^{43}$ erg s$^{-1}$;
7=Minimum random Lorentz factor of the injected particles;
8=Maximum random Lorentz factor of the injected particles;
9=Spectral slope of injected particles; 
10=Disk luminosity in units of $10^{47}$ erg s$^{-1}$; 
11=Radius of the BLR in units of $10^{17}$ cm. 



\clearpage


\vspace{-1.0cm}
\begin{figure}[h]
\begin{center}
\hbox{
\includegraphics[height=10cm,width=10cm]{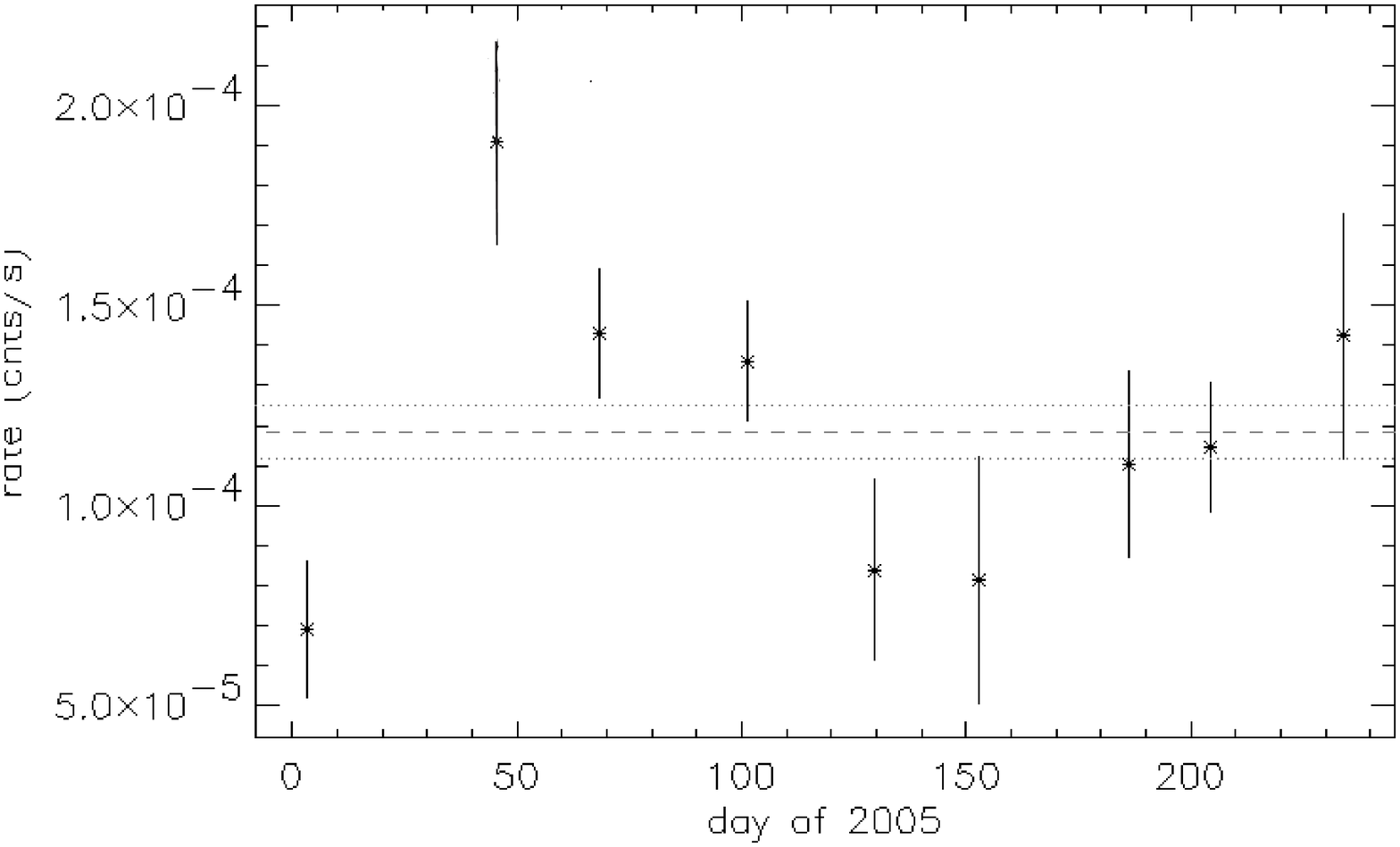}
}

\end{center}
\vspace{-1.0cm}
\caption{
{BAT light curve of 0836+710. The horizontal lines mark the BAT
average count rate. Significant variability of the hard X-ray flux is
detected on timescales of months. }}
\end{figure}

\clearpage


\vspace{-1.0cm}
\begin{figure}[h]
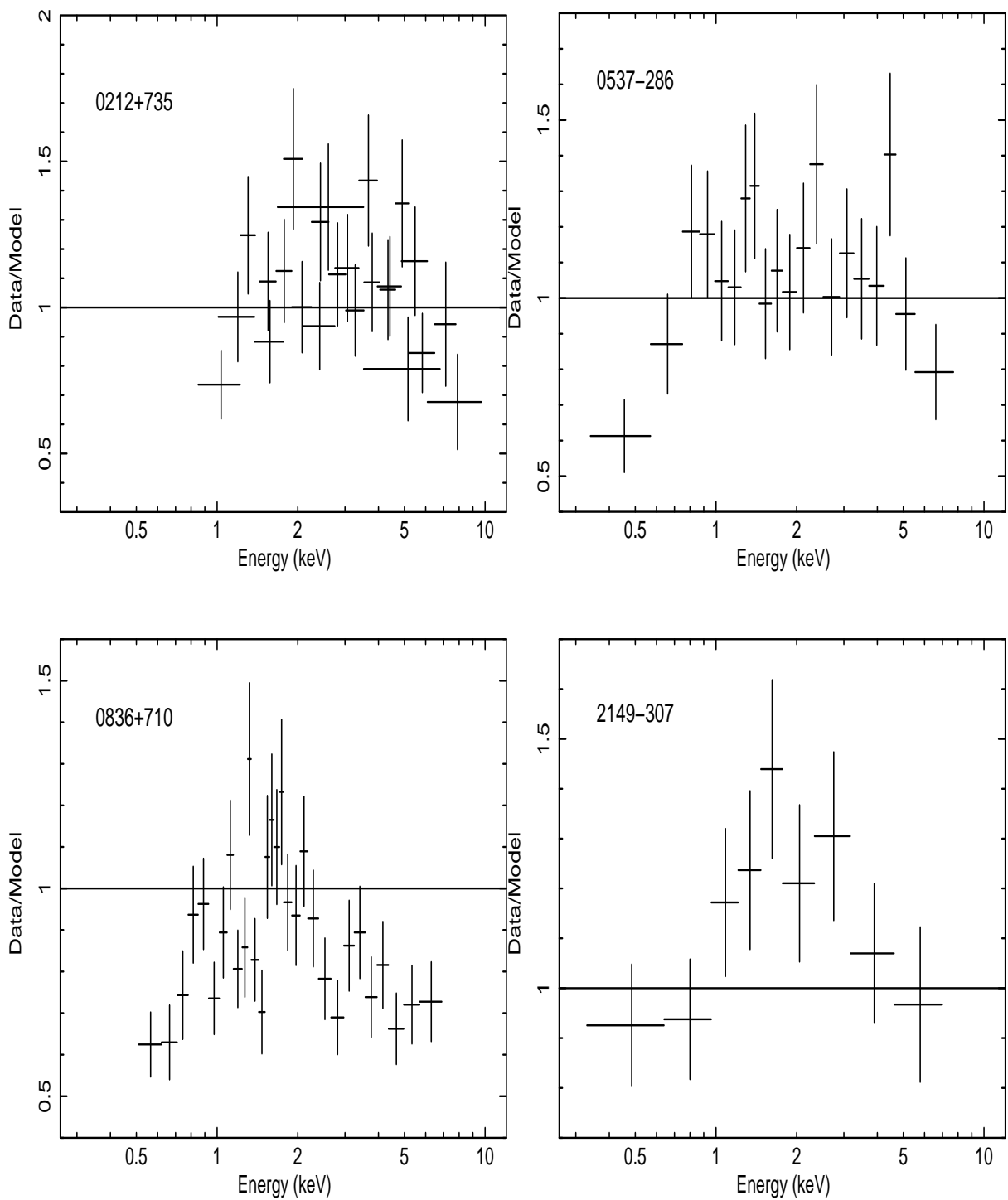

\begin{center}
\hbox{
\centerline{\includegraphics[height=8cm,width=9cm,angle=-90]{f2a.eps}\includegraphics[height=8cm,width=9cm,angle=-90]{f2b.eps}}
}
\end{center}

\hbox{
\centerline{\includegraphics[height=8cm,width=9cm,angle=-90]{f2c.eps}\includegraphics[height=8cm,width=9cm,angle=-90]{f2d.eps}}
}

\vspace{1.0cm}

\caption{
{Residuals of the joint fits to the XRT spectra of the four
quasars with a single power law and Galactic absorption. Spectral
flattening at low energies is present. For all sources except 0212+735
only the XRT data for the longest exposure are shown for clarity. 
}}
\end{figure}

\clearpage


\vspace{-1.0cm}
\begin{figure}[h]
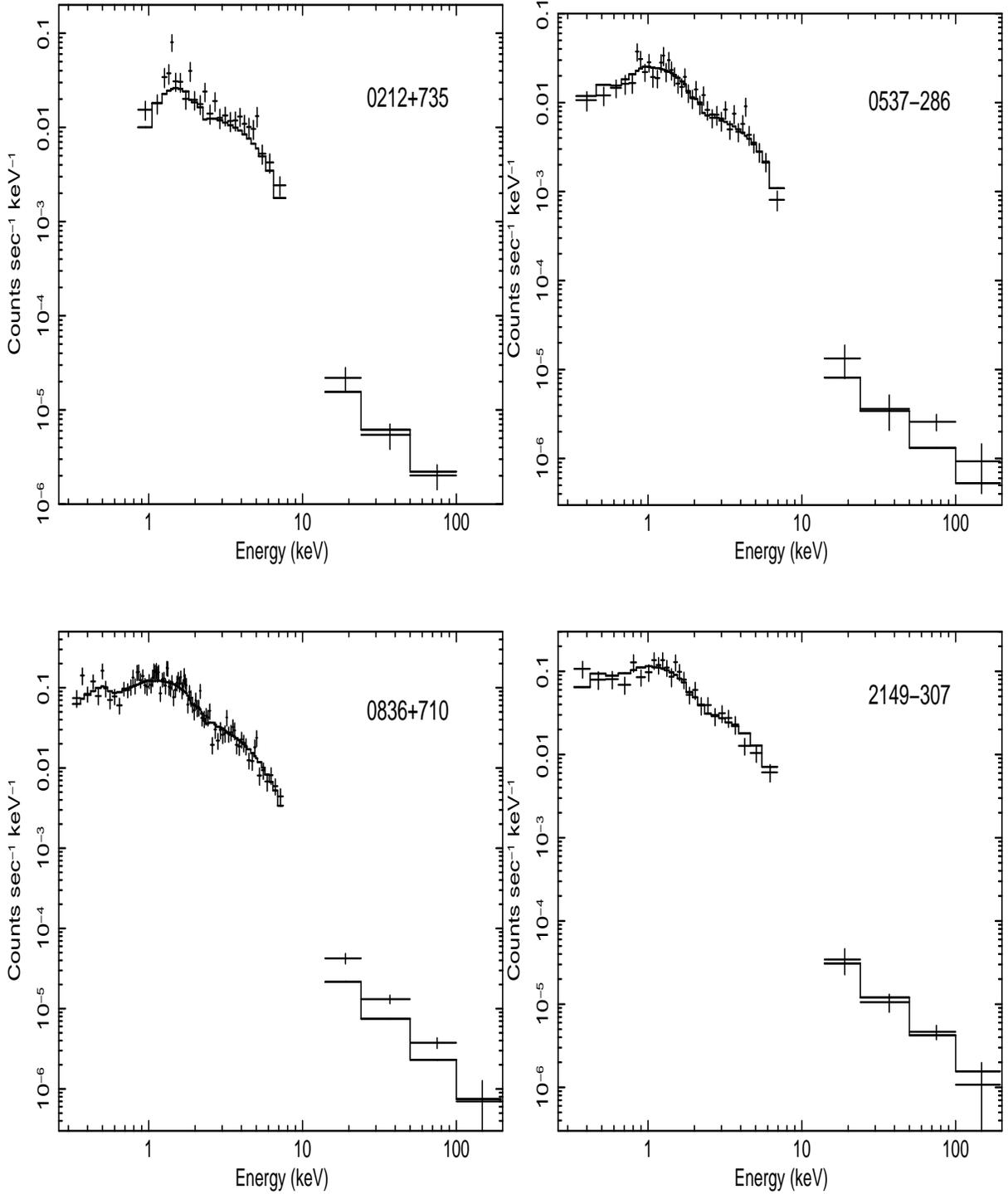

\begin{center}
\hbox{
\centerline{\includegraphics[height=8cm,width=9cm,angle=-90]{f3a.eps}\includegraphics[height=8cm,width=9cm,angle=-90]{f3b.eps}}
}
\end{center}

\hbox{
\centerline{\includegraphics[height=8cm,width=9cm,angle=-90]{f3c.eps}\includegraphics[height=8cm,width=9cm,angle=-90]{f3d.eps}}
}

\vspace{1.0cm}

\caption{
{Spectral fits to the joint XRT and BAT observations with the best-fit
models in Table~3c. Only the longest XRT exposure for each source is
shown for clarity. For 0836+710, note the large BAT residuals; the
latter are improve by the addition of a blackbody with $kT \sim 3$
keV, although the fit is not statistically better.  }}
\end{figure}

\clearpage 


\vspace{-1.0cm}
\begin{figure}[h]
\begin{center}
\hbox{
\centerline{\includegraphics[height=8cm,width=9cm]{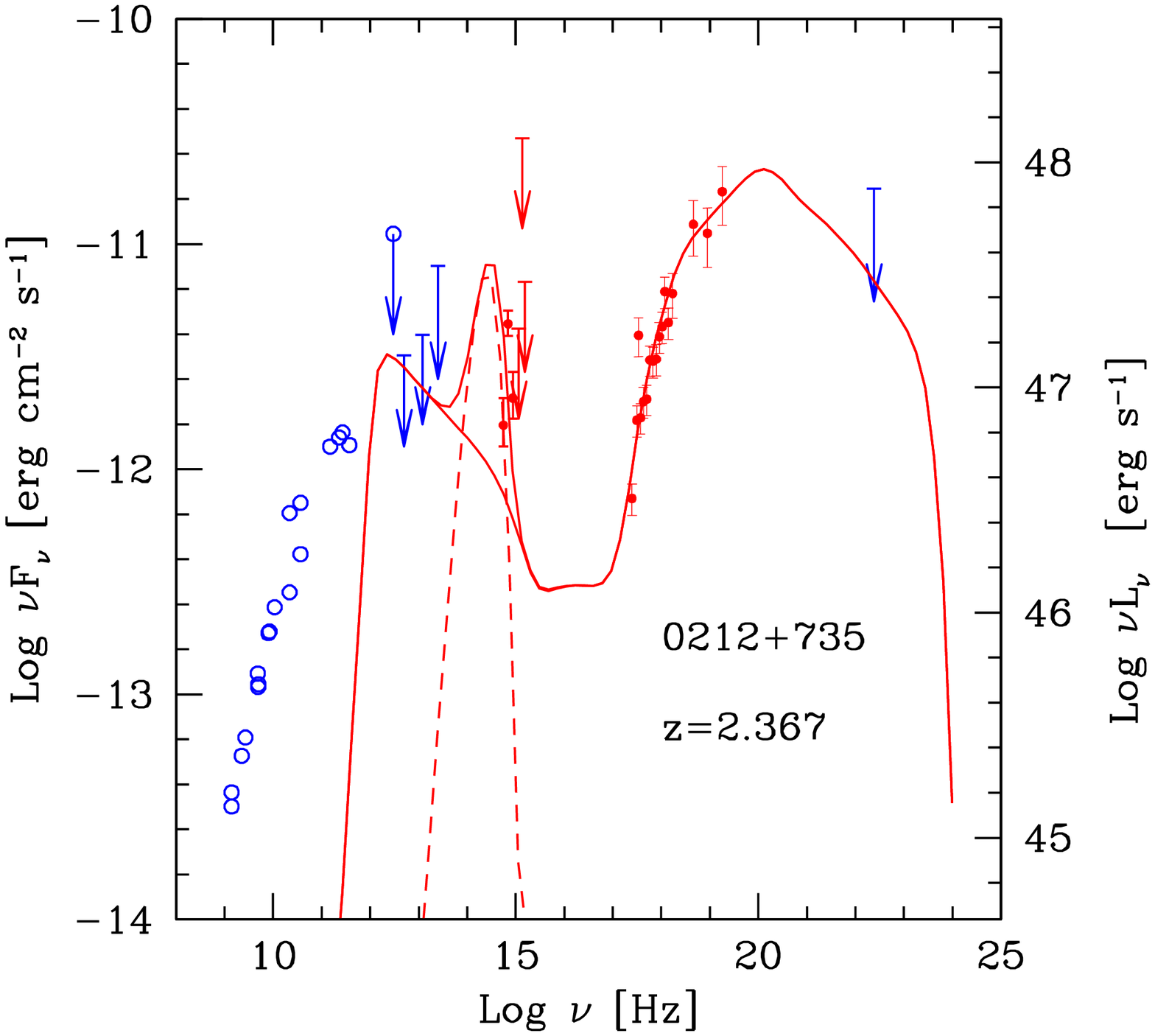}\includegraphics[height=8cm,width=9cm]{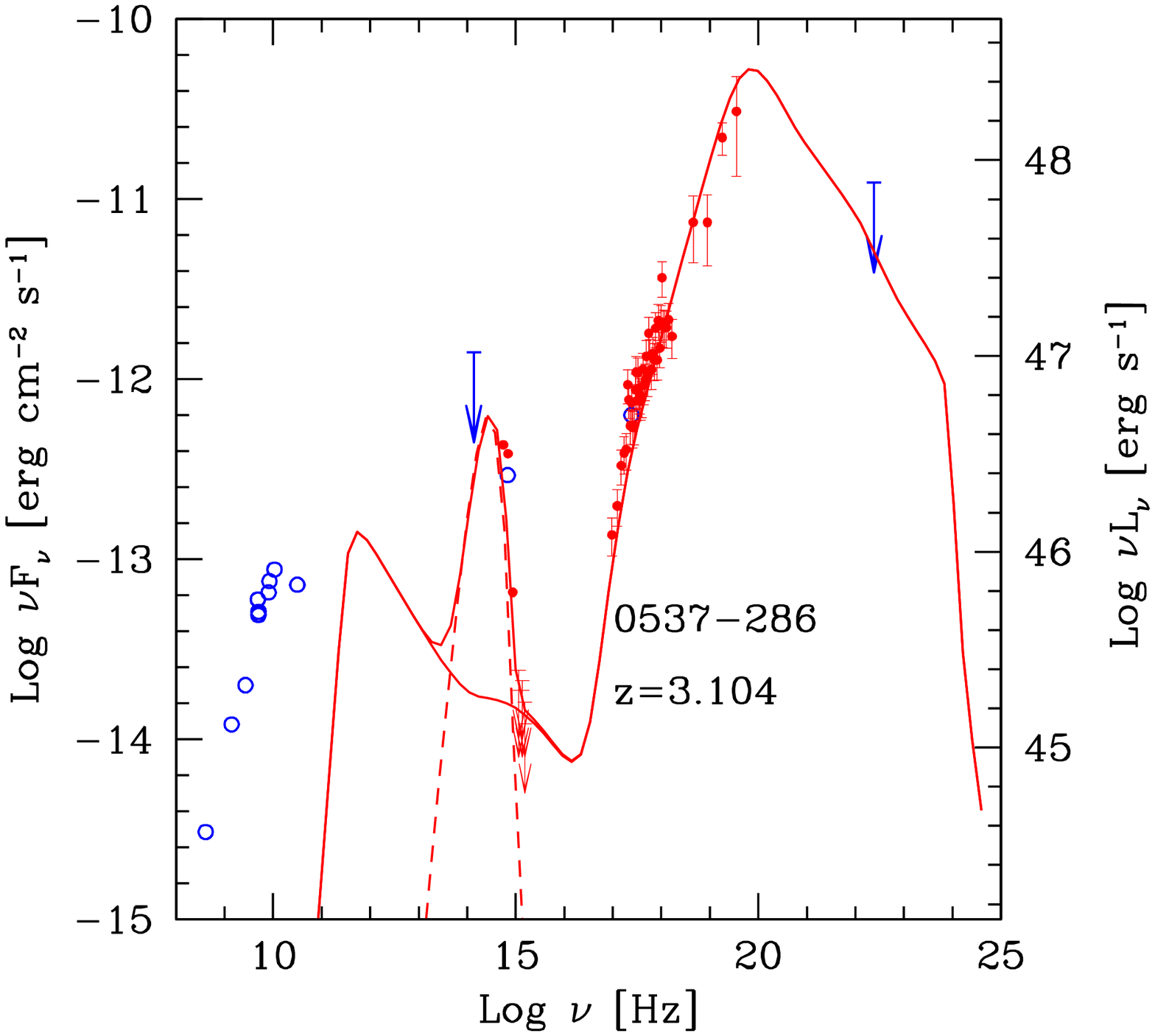}}
}
\end{center}

\hbox{
\centerline{\includegraphics[height=8cm,width=9cm]{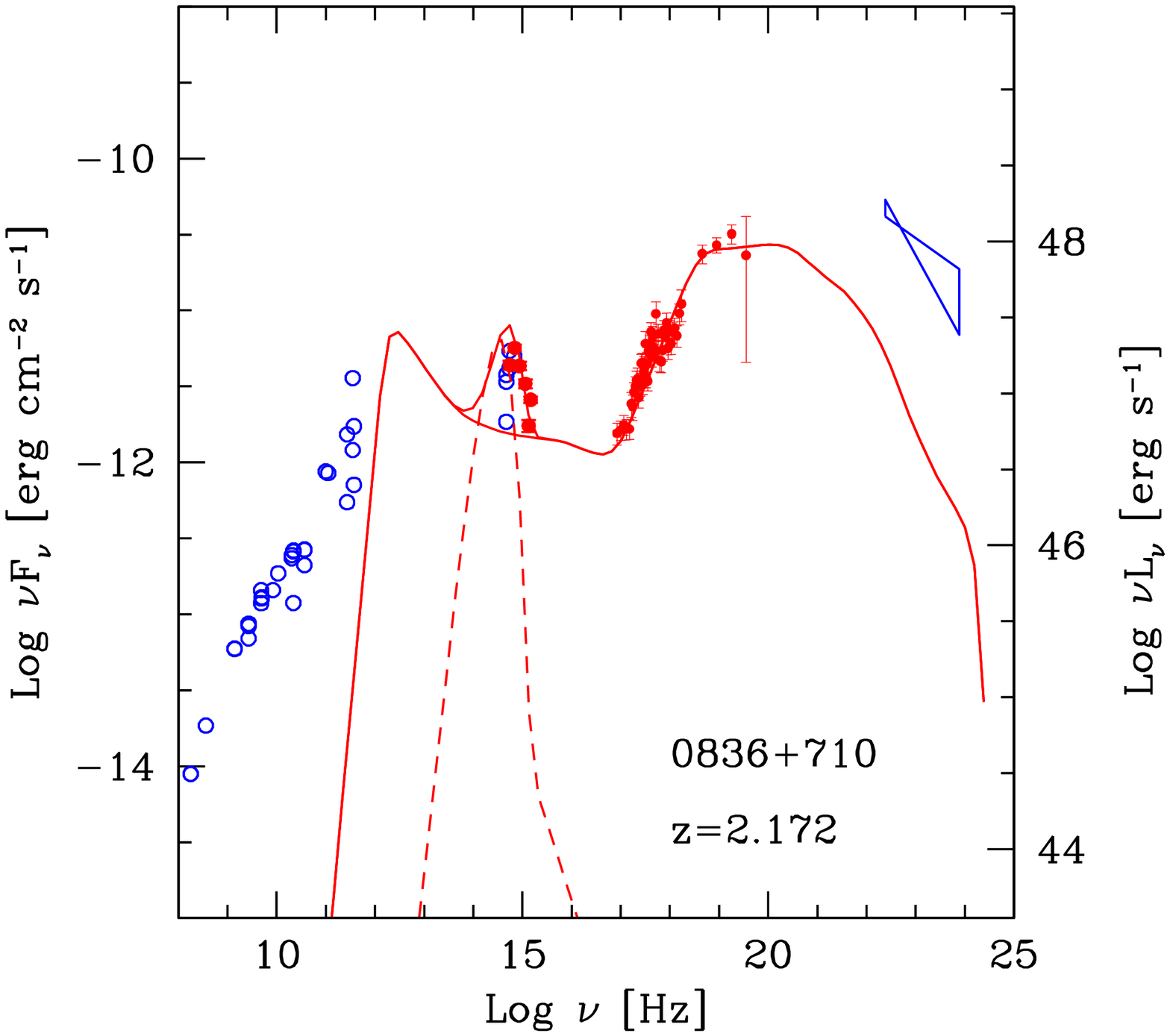}\includegraphics[height=8cm,width=9cm]{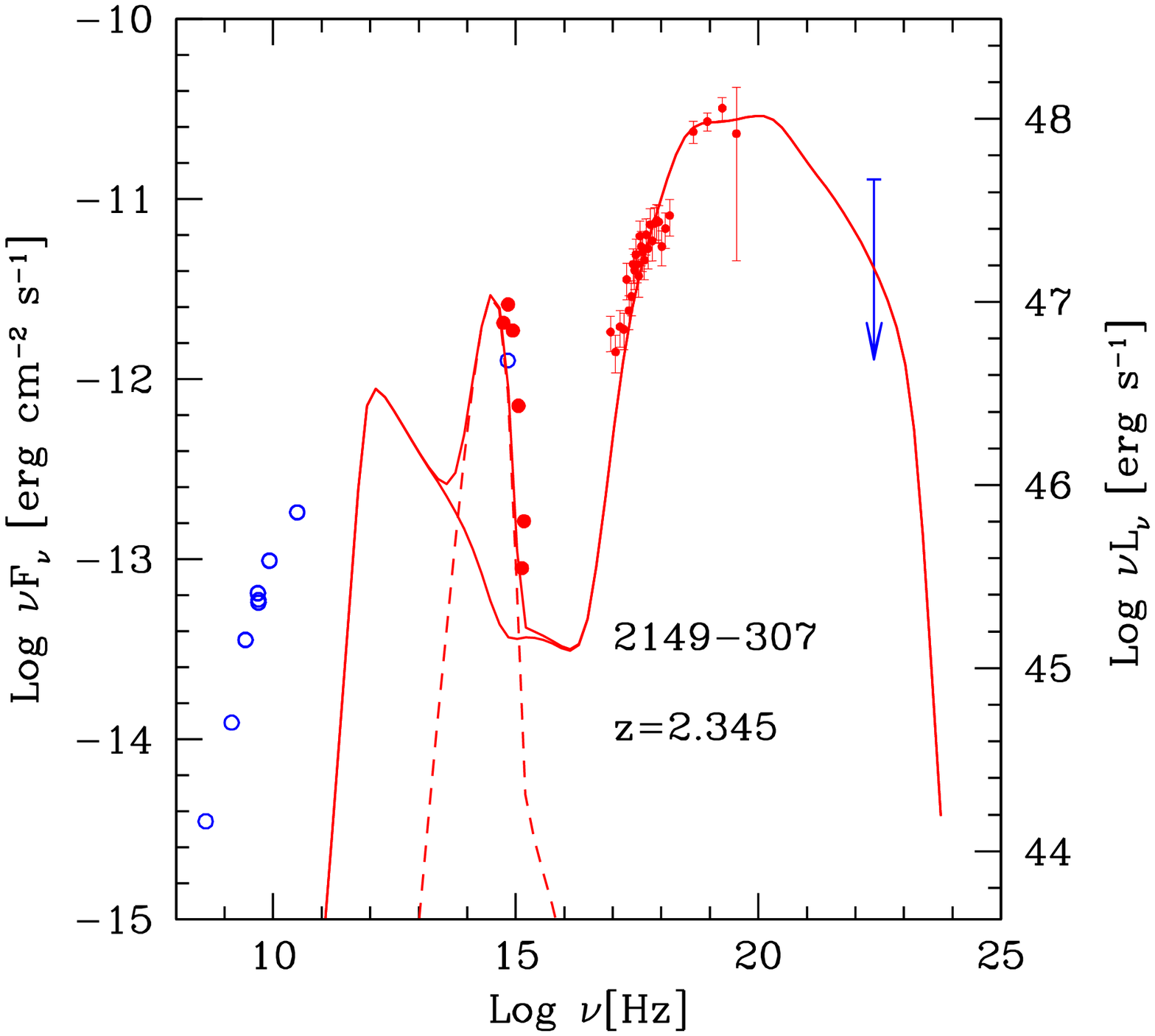}}
}

\vspace{1.0cm}

\caption{
{Spectral Energy Distributions of the four blazars. The
optical-to-X-ray data (filled dots) are from the UVOT and XRT, while
the BAT data represent the average over 9 months. The open dots are
archival data (Tavecchio et al. 2000). The solid lines represent the
best fit with a synchrotron+IC model, as in S06. The radio-through-IR
emission is due to synchrotron, the optical-UV to the disk and BLRs,
and the X-ray-to-MeV emission to Compton scattering, both of the
synchrotron (SSC) and external (EC) photons. }}
\end{figure}

\end{document}